\documentclass[fleqn,usenatbib]{mnras}
\usepackage{newtxtext,newtxmath}
\usepackage[T1]{fontenc}
\usepackage{ae,aecompl}
\usepackage{graphicx}	
\usepackage{amsmath}	
\usepackage{amssymb}	
\usepackage{xspace}
\usepackage{multirow}
\newcommand{\psr}{PSR~J1755\texorpdfstring{$-$}{-}2550\xspace}
\newcommand{\OurCode}{{\sc ComBinE}\xspace}
\usepackage{xcolor}

\title[Nature of the binary \psr]{\psr: A young radio pulsar with a massive, compact companion}

\author[C. Ng et al.]{
C. Ng$^{1,2,3}$\thanks{E-mail: cherry.ng@dunlap.utoronto.ca},
M.~U.~Kruckow$^{4}$,
T.~M.~Tauris$^{3,4}$,
A.~G.~Lyne$^{5}$,
P.~C.~C.~Freire$^{3}$,
\newauthor
A.~Ridolfi$^{3}$, 
I.~Caiazzo$^{1}$,
J.~Heyl$^{1}$,
M.~Kramer$^{3,5}$,
A.~D.~Cameron$^{3}$,
\newauthor
D.~J.~Champion$^{3}$,
B.~Stappers$^{5}$
\\
$^{1}$Dunlap Institute, University of Toronto, 
      50 St. George St., Toronto, ON M5S 3H4, Canada\\
$^{2}$Department of Physics and Astronomy,
      The University of British Columbia,
      Vancouver, BC V6T-1Z1, Canada\\
$^{3}$Max-Planck-Institut f\"{u}r Radioastronomie,
      Auf dem H\"{u}gel 69, D-53121 Bonn, Germany\\
$^{4}$Argelander-Institut f\"{u}r Astronomie, Universit\"{a}t Bonn,
      Auf dem H\"{u}gel 71, D-53121 Bonn, Germany\\
$^{5}$Jodrell Bank Centre for Astrophysics, 
      School of Physics and Astronomy, The University 
      of Manchester, Manchester M13 9PL, UK
}

\date{Accepted XXX. Received YYY; in original form ZZZ}

\pubyear{2018}

\begin{document}
\label{firstpage}
\pagerange{\pageref{firstpage}--\pageref{lastpage}}
\maketitle

\begin{abstract}
Radio pulsars found in binary systems with short orbital periods are usually fast spinning as a consequence of recycling via mass 
transfer from their companion stars; this process is also thought to decrease the magnetic field of the
neutron star being recycled. Here, we report on timing observations of the recently discovered binary \psr
and find that this pulsar is an exception: with a characteristic age of 2.1\,Myr, it is 
relatively young; furthermore, with a spin period of 315\,ms and a surface magnetic field strength at its poles
of $0.88 \times 10^{12}\,$G the pulsar shows no sign of having been recycled. 
Based on its timing and orbital characteristics, the pulsar either has a massive white dwarf (WD) or a neutron star (NS) companion. To distinguish between these two cases, 
we searched radio observations for a potential recycled pulsar companion and analysed archival optical data for a potential WD companion. Neither work returned conclusive detections. 
We apply population synthesis modelling and find that both solutions are roughly equally probable. Our population synthesis also predicts a minimum mass of 0.90\,$M_{\odot}$ for the companion star to \psr\ and we simulate the systemic runaway velocities for the resulting WDNS systems which may merge and possibly produce Ca-rich supernovae.
Whether \psr hosts a WD or a NS companion star, it is certainly a member of a rare subpopulation of binary radio pulsars.
\end{abstract}


\section{INTRODUCTION} \label{sec:intro}
\begin{figure}
\includegraphics[width=3.2in]{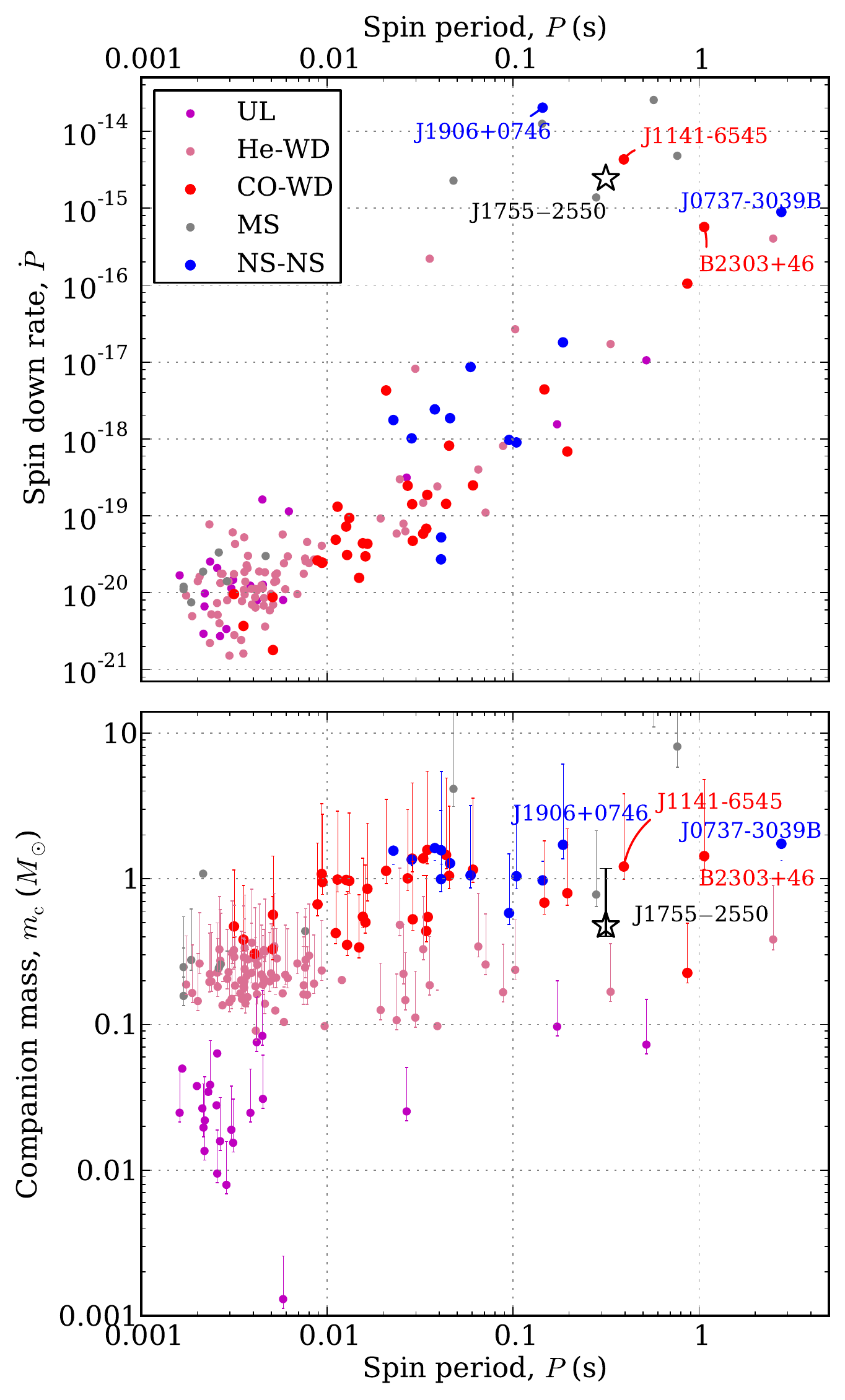}
\caption{Top panel: The classical $P-\dot{P}$ diagram  of all binary radio pulsars. Plotted here are the observed $\dot{P}$ values without any correction for the kinematic Shklovskii effect \citep{shk70}.
Bottom panel: Companion mass versus spin period of all binary pulsars. For systems with unknown orbital inclination we plot the median companion mass instead, corresponding to $i=60^{\circ}$. The error bars indicate the range of minimum to maximum companion mass corresponding to $i=90^{\circ}$ and $i=26^{\circ}$, respectively. In both panels, the data points are colour-coded according to their binary companion type. Ultra-light (UL) companions are represented by black, helium (He) WDs by pink, carbon-oxygen (CO) WDs by red, main-sequence stars by gray and DNS (NS-NS) systems by blue. The subject of this paper, \psr, is shown as a star symbol. The four binary systems where the observed pulsar is the second-formed compact object are also annotated.}
\label{fig:MSP}
\end{figure}

The All-Sky High Time Resolution Universe (HTRU) Pulsar survey \citep{kjv+10} conducted with the 64-m Parkes radio telescope between 2010 and 2015 has greatly increased the number of known pulsars in binary systems. Among these new discoveries is \psr \citep{ncb+15}. At the time of the publication of the discovery paper, only a preliminary timing solution was available for this pulsar:
it was known to have a relatively long spin period ($P=315\;{\rm ms}$), an orbital period of 9.7\,d
and an orbital eccentricity of about 0.09.
Further timing campaigns at the Lovell and the Effelsberg telescopes spanning 2.6\,yr have significantly improved the positional uncertainty and broken its degeneracy with the spin-down rate. We find in this work that the pulsar has a large spin-down rate ($\dot{P}\, = \, 2.4 \times\, 10^{-15}$). These values place \psr in a region of the $P-\dot{P}$ diagram that, although densely populated by normal, non-recycled, isolated pulsars, is very sparsely populated by binaries (see Fig.~\ref{fig:MSP}). 
 
The only way for the binary pulsar \psr to avoid having been recycled is if either it has a main-sequence (MS) star companion or it is the second-formed compact object in the system.
As we shall argue later (see Section~\ref{subsec:MS-companion}), we find that the scenario with a MS companion is the least likely. Therefore, in this paper, we focus our investigation on whether \psr could be the second-formed compact object in a binary. Such a binary is very unusual; in fact there are only four known cases, which are annotated in Fig.~\ref{fig:MSP} for comparison. PSRs~B2303+46 
\citep{tc99} and J1141$-$6545 \citep{mlc+00} are the only two binaries where a young neutron star (NS) is orbiting an old, massive white dwarf (WD) companion, confirmed by their optical identifications \citep{kk99,abw+11}. This requires a fine-tuned formation scenario with an initial binary of two stars with typical masses in the range $6-10\;M_{\odot}$, which undergo mass reversal during the mass-transfer phase \citep{ts00,drk02}. Another possible member of this population is PSR~J1906+0746 \citep{lsf+06,vks+15}, however, the nature of its companion has not yet been confirmed, but given its mass
($1.322 \, \pm \, 0.011\, M_{\odot}$) it is likely to be a NS. That would imply PSR~J1906+0746 is the second-formed NS in a double neutron star system (DNS), with the first-formed likely to be a (still
undetected) recycled pulsar. Such a case is again a statistically rare find, because in a DNS the second-formed (non-recycled) NS has a much shorter radio lifetime compared to the recycled, first-formed NS. Indeed, out of all the known DNS systems, it is almost always the first-formed NS that is observed as a radio pulsar. The only confirmed  case where we see the second formed pulsar is PSR~J0737$-$3039B, but in this system we also see the first-formed pulsar, PSR~J0737$-$3039A: this is the well-known ``double pulsar'' system \citep{bdp+03,lbk+04}.

In either case, \psr has an unusual formation history and is thus an object of interest for binary stellar evolution. In Section~\ref{sec:obs} of this paper, we describe the radio observations taken and present an update of the timing solution. In Section~\ref{sec:Formation} we discuss various possible formation scenarios of \psr, and in Section~\ref{sec:popsyn} we present a population synthesis investigation for WDNS binaries. Further potentially observable clues are discussed in Section~\ref{sec:discussions}, and we conclude our findings in Section~\ref{sec:conclusion}.

\section{RADIO TIMING OBSERVATIONS} \label{sec:obs}

\subsection{Observational set-up}
The majority of the timing observations of \psr have been taken at the Jodrell Bank Observatory with the Lovell 76-m telescope, using a Digital Filterbank system (DFB) backend. The DFB is based on the implementation of a polyphase filter in FPGA processors with incoherent dedispersion. 
The Jodrell DFB has a bandwidth of 384\,MHz with a central frequency at 1532\,MHz. 
These observations have roughly weekly cadence and each integration is of the order of half an hour.
A handful of DFB observations were also recorded at Parkes, with a bandwidth of 256\,MHz and a central frequency of 1369\,MHz. 

A dedicated timing campaign was conducted at the Effelsberg 100-m radio telescope, with the main goal of obtaining high-quality multi-frequency polarimetry data and to perform a deep search for a potential neutron star companion of \psr. All Effelsberg observations were made using the PSRIX backend \citep{Lazarus+16} in baseband mode with a nominal bandwidth of 200\,MHz. 
About 5\,hr were spent all together observing at a central frequency of 4.8\,GHz (C-band). No detection is made above a signal-to-noise ratio of 5. 

\subsection{Derivation of times-of-arrival and the timing solution}
Our analysis of the radio timing data made use of the \textsc{psrchive} data analysis package \citep{hvm04}.
Each observation is first corrected for dispersion and folded at the predicted topocentric pulse period. 
A high signal-to-noise template is created by 
co-adding all available observations. 
This template is then convolved with each individual profile to produce a time-of-arrival (TOA) \citep{tay92}. 
We generate one TOA per observation by scrunching in time and frequency to maximize the signal-to-noise ratio of each TOA. The spin and orbital periods of \psr are relatively long and the high DM of 751\,cm$^{-3}\,$pc means little scintillation. Hence no timing parameter should vary significantly over the course of each of the half hour integration.
The DE421 Solar System ephemeris of the Jet Propulsion Laboratory \citep{fwb09} is used to transform the TOAs to the Solar System barycentre. 
\begin{figure}
\centering
\includegraphics[width=3.3in]{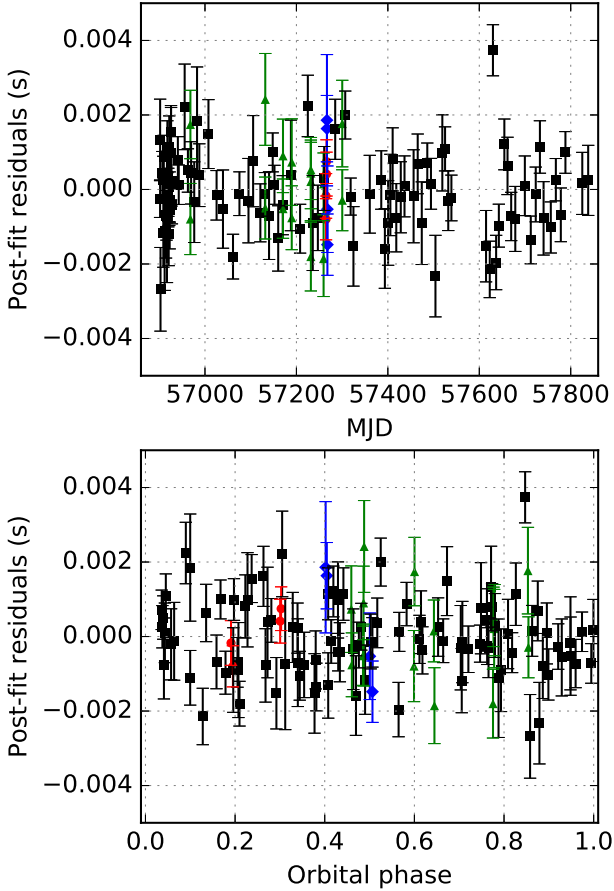}
\caption{Post-fit timing residuals of \psr with the parameters listed in Table.~\ref{tab:timing} taken into account. We use different colours to represent different data sets, with black squares being Jodrell 1.3\,GHz TOAs, green triangles being Parkes 1.3\,GHz TOAs, red circles being Effelsberg 1.3\,GHz and blue diamonds being Effelsberg 2.6\,GHz TOAs. The top panel shows residuals as a function of MJD, whereas the bottom panel plots residuals as a function of orbital phase.}\label{fig:toas}
\end{figure}

\begin{table}
    \centering
  \caption{Best-fit parameters for \psr.  Values in parentheses are the nominal 1-$\sigma$ uncertainties in the last digits.}
 \begin{minipage}{9cm}
\begin{tabular}{p{5.7cm}p{2.4cm}}
\hline
\multicolumn{2}{c}{Spin and astrometric parameters} \\
\hline
Right ascension, $\alpha$ (J2000) &  17:55:38.400(4) \\
Declination, $\delta$ (J2000) & $-$25:50:22.0(18) \\
Spin period, $P$ (ms) & 315.1960620987(16)\\
Period derivative, $\dot{P}$ & 2.4337(14)$\times10^{-15}$\\
Dispersion measure, DM (cm$^{-3}\,$pc) & 750.9(4) \\
\hline
\multicolumn{2}{c}{Binary parameters}\\
\hline
Orbital period, $P_{\rm{orb}}$ (days) & 9.6963342(6)\\
Projected semi-major axis, $x$ (lt-s) & 12.28441(14) \\
Epoch of periastron, $T_{\rm{0}}$  (MJD) & 56904.1265(4) \\
Eccentricity, $e$ & 0.08935(2)\\
Longitude of periastron, $\omega$ ($^{\circ}$) &  129.680(15) \\
\hline
\multicolumn{2}{c}{Timing model} \\
\hline
Timing epoch (MJD) & 57329 \\
First TOA (MJD) &  56901.8 \\
Last TOA (MJD) & 57848.3  \\
Weighted RMS residuals (ms) & 0.99 \\
Reduced $\chi^{2}$ $\ddagger$ & 1.9 \\
Solar System ephemeris & DE421 \\
Binary model & DD \\
\hline
\multicolumn{2}{c}{Derived parameters}\\
\hline
DM-derived distance (kpc) & 4.91$-$10.29$\dagger$  \\
Mean flux density at 1.3\,GHz, $S_{1.3\,\mathrm{GHz}}$ (mJy) & 0.20 \\
Mean flux density at 2.6\,GHz, $S_{2.6\,\mathrm{GHz}}$ (mJy) & 0.04 \\
Characteristic age, $\tau$ (Myr) & 2.1 \\
Characteristic dipole surface magnetic & \\
field strength at equator, $B_{\rm{eq}}$ ($10^{12}$ G) & 0.88 \\
Mass function ($M_{\sun}$) & 0.0211707(16)\\
Minimum companion mass$^{*}$, $m_{\rm{c,min}}$ ($M_{\sun}$) & 0.39\\
Median companion mass$^{**}$, $m_{\rm{c,med}}$ ($M_{\sun}$) & 0.47 \\
\hline \label{tab:timing}
 \end{tabular}
\vspace{-0.5\skip\footins}
 \begin{flushleft}
  $^{\ddagger}$ The reduced $\chi^{2}$ stated here represents the value before the application of EFAC. Note that the rest of the timing solutions have EFACs incorporated, bringing the reduced $\chi^{2}$ to unity. \\
$^{\dagger}$ Using the electron density model from \citet{YMW16} we obtain a smaller derived distance of 4.9\,kpc while the \citet{cl02}  model predicts a further distance of 10.3\,kpc.\\
 $^{*}$ $m_{\rm{c,min}}$ is calculated for an orbital inclination of $i=90^{\circ}$ and an assumed pulsar mass of $1.3\,M_{\sun}$. \\
 $^{**}$ $m_{\rm{c,med}}$ is calculated for an orbital inclination of $i=60^{\circ}$ and an assumed pulsar mass of $1.3\,M_{\sun}$. \\
 \end{flushleft}
\end{minipage}
\end{table}

Finally, the \textsc{tempo2} software package \citep{hem06} is used to fit a 
timing model to all TOAs, taking into account the astrometry, spin, and orbital motion of the pulsar. 
To describe the orbit of \psr we have used the
Damour-Deruelle (DD) timing model \citep{dd86} in \textsc{tempo2}; this is a theory-independent description of eccentric binary orbits.
Since almost all of our timing data were taken at 1.3\,GHz with a narrow bandwidth, we do not have a good handle on the precision of the DM and hence have fixed the DM at the nominal value of 751\,cm$^{-3}\,$pc.  
The time span of our data set is still too short to constrain any proper motion and parallax. Given the limited timing precision typically associated with a slow pulsar, it might be years before we can reliably measure proper motion and parallax. We have held these parameters fixed at zero. 
As a last step, we compensate for any remaining systematic effects (e.g. instrumental or minor radio frequency interference) by calculating dataset-specific calibration coefficients (also known as `EFAC'). These coefficients are applied to scale the TOA uncertainties such that each final respective reduced $\rm\chi^2$ is unity, in order to produce reliable uncertainty in the fitted parameters. See Fig.~\ref{fig:toas} for the post-fit timing residuals.
If we assume a pulsar mass of 1.3\,$M_{\odot}$, using Equation~(2) of \citet{wt81} we obtain a predicted precession of periastron ($\dot\omega$) in general relativity of 0$_{.}^{\circ}0069$\,yr$^{-1}$. Given our current precision in the measurement of $\omega$ (0$_{.}^{\circ}015$, see Table~\ref{tab:timing}), we do not expect to be able to detect $\dot\omega$ in the short term.

\subsection{Pulse profile and polarization study} \label{sec:profile}

\begin{figure}
\centering
\includegraphics[width=3.3in]{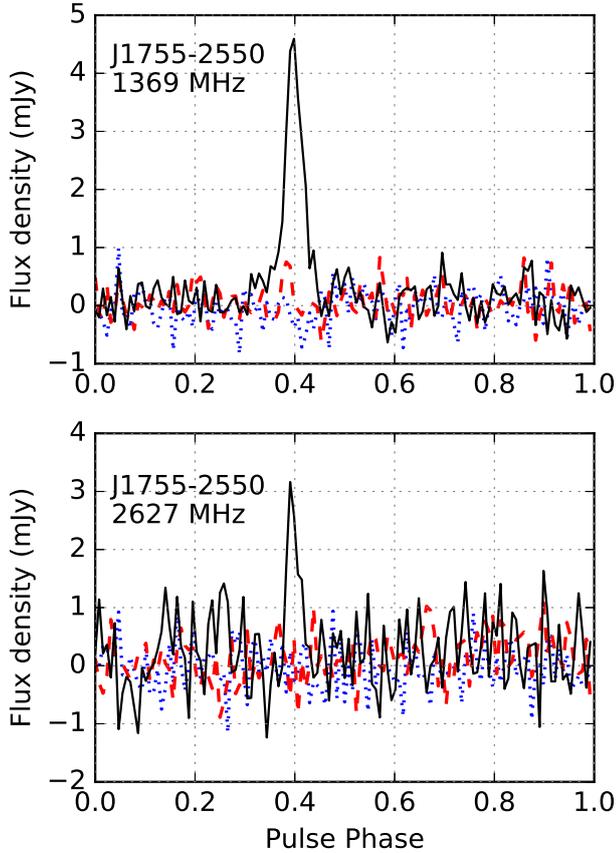}
\caption{
Pulse profiles observed at Parkes at 1.3\,GHz (top panel) and at Effelsberg at 2.6\,GHz (bottom panel). 
The integrated profiles shown here have 128 phase bins. The total intensity is represented by the black solid line, linear 
polarization by the red dashed line and circular polarization by the blue dotted line.}
\label{fig:pol}
\end{figure}

The Parkes and Effelsberg observations record four Stokes parameters in each frequency channel and 
thus can be used to study the polarization profile. 
A calibration scan was taken before or after each of the targeted pulsar observations. 
This calibration scan triggers a square-wave signal of the noise diode coupled to the receptors in the feeds, 
which can be used to polarization calibrate for the differential gain and phase between the feeds, in turn enabling the retrieval of the true Stokes parameters.

Fig.~\ref{fig:pol} show the integrated polarization profiles of
\psr in total intensity, linear and circular polarization, observed at 1.3 and 2.6\,GHz respectively. No significant position angle (PA) has been measured. \psr has a simple profile comprising only one component and does not appear to be polarized linearly nor circularly. We also do not obtain any constraining rotation measure. It has been proposed in the literature that young, energetic pulsars with $\dot{E}>10^{35}\;{\rm erg\,s}^{-1}$ tend to have significant linear polarization \citep[see, for example, Fig. 8 in][]{wj08b}.
\psr has a relatively high $\dot{E}$ of the order of $10^{33}\;{\rm erg\,s}^{-1}$. The lack of linear polarization in this case is note-worthy but not inconsistent with literature. 
We convolve the profile at 2.6\,GHz with a scattering tail, and measured a characteristic scattering timescale of  $\sim$13.5(14)\,ms at 1\,GHz. This small hint of scattering could also have decreased the amount of polarization.

\section{Formation scenario} \label{sec:Formation}
\subsection{Formation of binary pulsars}
The standard formation scenario of a pulsar in a binary system is reasonably well established in the literature \citep[see e.g.][]{bv91}. 
It all begins with two main-sequence stars, where the initially more massive star evolves first, expands and transfers mass to its companion star, before it undergoes a supernova (SN) explosion to produce a NS. The newborn NS gradually spins down afterwards, as it radiates its rotational energy similar to the case of a normal radio pulsar \citep{lk04}. At a later stage the secondary star expands after depletion of hydrogen core burning and initiates mass transfer to the NS. In this process, known as ``pulsar recycling'', the NS becomes rejuvenated as it gains mass and angular momentum \citep[e.g.][]{acrs82,tv06}.
At the same time, the strength of its surface magnetic field is reduced \citep[e.g.][]{bha02}. Hence, the outcome of the recycling process is an old NS with rapid spin (small value of $P$), which enables the radio emission mechanism to re-activate, and a small $\dot{P}$ as a result of the reduced B-field. 

From the $P-\dot{P}$ diagram in the top panel of Fig.~\ref{fig:MSP}, it can be seen that most of the binary pulsars cluster around short spin periods of millisecond duration (i.e. MSPs) and small spin-down rates of the order of $10^{-21}$ to $10^{-19}$. 
The longer the duration of this recycling phase, during which the source is visible as an X-ray binary, the faster the final spin period of the pulsar and the smaller the period derivative. An important factor determining the degree of recycling is therefore the initial mass of the secondary star \citep[see e.g.][for a review]{tau11}. The more massive the secondary star is, the less efficient is the recycling process.

If the secondary star is sufficiently massive, it will undergo a SN explosion itself to form a younger, second NS. There are about a dozen or so of these DNS binaries known to-date \citep[e.g.][]{msf+15,lfa+16}. 
As can be seen in Fig.~\ref{fig:MSP}, the first-formed NSs in DNS systems tend to have relatively long spin periods and large period derivatives compared to MSPs. These NSs are therefore observed as mildly recycled pulsars and their properties are brought about by the so-called Case~BB Roche-lobe overflow \citep{tlp15}, following the high-mass X-ray binary and common-envelope phase \citep{tv06}.

Finally, if the secondary star is not massive enough to undergo core collapse, the mass-transfer phase can last much longer \citep[up to several Gyr,][]{ts99}, before the companion star eventually sheds its outer layer and results in a white dwarf (WD). This often leads to very efficient recycling.
As apparent from Fig.~\ref{fig:MSP} \citep[see also Fig.~9 in][]{tlk12}, the most recycled systems are indeed those with ultra-light (UL) and helium WD (He~WD) companions, followed by those with heavier masses, namely carbon-oxygen (CO~WD) or oxygen-neon-magnesium (ONeMg~WD) WDs \citep{ltk+14}. The NSs displaying the least amount of recycling are those found in DNS systems.

\subsection{\psr\--- the case of a young pulsar}
The combination of a large $\dot{P}$ value ($2.4\times10^{-15}$) and a relatively slow spin period ($P=315\;{\rm ms}$)
identifies \psr as being a non-recycled radio pulsar (i.e. there are no signs of accretion onto the NS after its formation). 
The pulsar is thus relatively young and a member of a binary system with a
companion star in the mass range $m_{\rm{c}}=0.4-2.0\;M_{\odot}$ at the 95\% C.L. (based on its measured mass function, see Section~\ref{subsec:mass-function}). The possibilities of the nature of the companion star, seem therefore to be restricted to the following three possibilities: a MS star, a WD or a NS.
We now investigate each of these cases in more detail.

\begin{figure}
\includegraphics[width=3.2in]{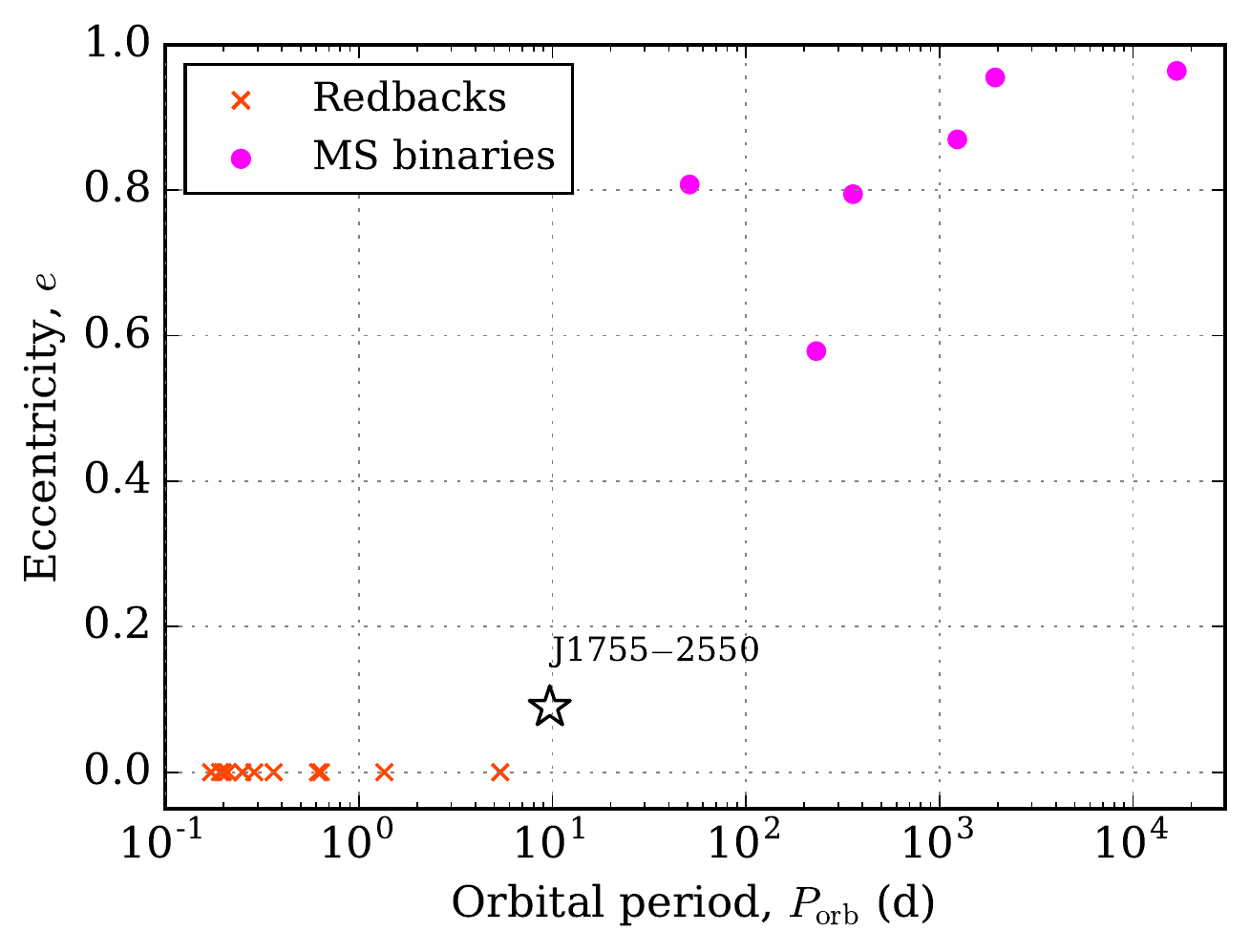}
\caption{Eccentricity as a function of orbital period for binary radio pulsars with hydrogen-rich companions in the Galactic disk. These systems can host either so-called redbacks (stripped MS stars) or regular MS stars.  \psr is plotted for comparison and does not fit in either of these two populations, see Section~\ref{subsec:MS-companion}. The error bars on all data points are much smaller than the size of the plotted dots.}
\label{fig:eccentricity}
\end{figure}

\subsection{A MS star companion?}\label{subsec:MS-companion}
In Fig.~\ref{fig:eccentricity}, we have plotted the eccentricities as a function of orbital period for all Galactic disk binary radio pulsars with hydrogen-rich companion stars. These companions can be either so-called redbacks (stripped MS stars) or regular MS stars. The redback systems \citep{rob11,ccth13} are all fast-spinning MSPs with low B-fields in circular orbits. Hence, we can rule out the latter possibility for \psr. 

We note that \psr has a much smaller eccentricity ($e\simeq 0.09$) than all the known young radio pulsars with MS-star companions which have $e\ga 0.60$ (see Fig.~\ref{fig:eccentricity}). 
Furthermore, of the six known radio pulsars with a MS-star companion only one system has a MS star with a mass which is potentially less than $3\;M_{\odot}$ \citep[PSR~B1820$-$11,][]{cl86,lm89} and that system has an orbital period of 358~days (i.e. much larger than that of \psr). It has even been suggested \citep{py99} that this companion star might be a WD.
Taken together, all these strongly suggests that the companion of \psr is not a regular MS star, although this possibility cannot be ruled out completely (see also Section~\ref{subsec:popsyn_MS} for further discussions based on population synthesis).

However, as we will see in 
Section~\ref{sec:discussions}, there is no  clear evidence for any MS star associated with \psr\, in
optical observations. Furthermore, there is no evidence for orbital variability (see bottom panel of Fig.~\ref{fig:toas}) normally associated
with tidal and rotational effects caused by an extended object, as observed
for the pulsars with optically identified MS companions. We do not measure any variations in the orbital period ($\dot{P}_{\rm{orb}}$), with a statistically insignificant best-fit value of $-2(5)\times10^{-9}$. Nor do we see any evidence for any eclipses
that might have been caused by outgassing of such a companion, which are also common observations
in systems with identified MS companions.

In the rest of this paper, we therefore investigate the more likely case of \psr being the last-formed member of a double degenerate system. The interesting question now is whether it has a WD or a NS companion star (i.e. whether it is a WDNS system, where the NS formed {\it after} the WD, or a DNS system). 
In the following, we discuss the two different possibilities based on the mass function, its orbital parameters, and the outcome of a population synthesis simulation.

\subsection{Mass function of \psr}\label{subsec:mass-function}
\begin{figure}
\includegraphics[width=3.3in]{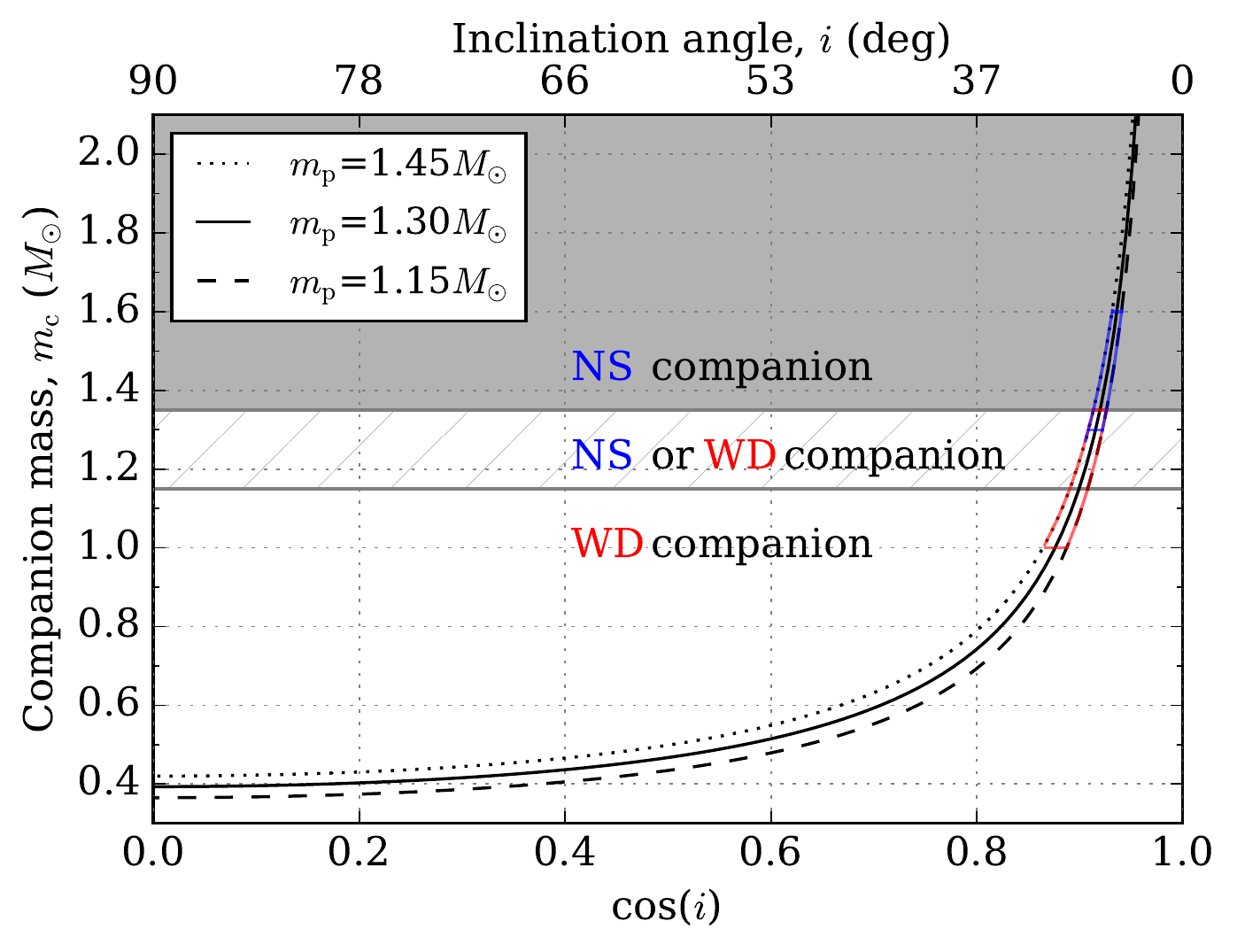}
\caption{Companion mass ($m_\mathrm{c}$) as a function of $\cos (i)$. Here $i$ the orbital inclination angle and $\cos (i)$ is a quantity with uniform distribution for randomly oriented orbits
of the \psr system. The different curves correspond to the unknown mass of \psr and represent typical masses of the second-formed NS in DNS systems. The grey-shaded area represents roughly the region where \psr has a NS companion. The semi-shaded area marks the region where \psr could have either a NS or a WD companion. The most likely companion mass ranges for NS (blue) and WD (red) are indicated in accordance with the discussion in Section~\ref{subsec:mass-function}. An orbital inclination of $\le 26^{\circ}$ is required for a DNS system.}
\label{fig:mass-function}
\end{figure}

In Fig.~\ref{fig:mass-function}, we have plotted the companion star mass as a function of the unknown orbital inclination angle.
At first sight, the somewhat small mass function of \psr  ($f=0.0212\;M_{\sun}$) strongly favors a WDNS system compared to a DNS system. Based on this function alone, and assuming a pulsar mass of $1.30\;M_{\sun}$ \citep[a typical NS mass of a young, binary, non-recycled pulsar, e.g.][]{tkf+17}, there is only about a 5--6\% chance that \psr has a NS companion, requiring an almost face-on orbit with an orbital inclination angle between $26^{\circ}-17^{\circ}$ for an assumed first-born NS mass between $1.15-2.1\;M_{\sun}$.

However, our population synthesis (see Section~\ref{subsec:Porb_ecc}) predicts a minimum WD mass of $0.90\;M_{\sun}$ (Fig.~\ref{fig:heatmap+histogram}) for any WD member of a WDNS system. 
This is not surprising since to produce such a system a mass reversal between the two stars is needed, such that the WD forms first (from the originally most massive star, i.e. the primary star) and the NS forms afterwards from the secondary star which accretes enough mass from the (giant) primary star that it undergoes core collapse to produce a NS \citep[e.g.][]{ts00}. Hence, a WDNS system can form from an initial zero-age main sequence (ZAMS) binary with two relatively massive stars near the threshold limit for producing a NS. As an illustrative example, consider a ZAMS binary with $9\;M_{\sun}$ and $8\;M_{\sun}$ stars. The $9\;M_{\sun}$ primary star is slightly too light to undergo core collapse, thus it necessarily forms a massive ($> \, 0.90\;M_{\sun}$) WD. However, the $8\;M_{\sun}$ secondary star accretes enough material to go over the threshold ($\sim 10-11\;M_{\sun}$) to
produce a NS. When it goes SN it produces a young pulsar, which never gets recycled.

Given that WDNS systems only form with a massive WD ($>\,0.90\;M_{\sun}$), and that NSs also have a mass above this limit, means that the first-formed compact object (i.e. the current companion star to the observed radio pulsar) must have a mass of at least $0.90\;M_{\sun}$. This fact, in combination with the measured mass function, means that the relative statistical probability for \psr being a DNS system is much greater than the aforementioned 5-6\,\%.

As a first na\"{i}ve guess, one could assume a first-born NS with a mass between $1.15-2.1\;M_{\sun}$. In this case, we find that the probability for \psr being a DNS system is about 46\%. This is estimated assuming an {\em a~priori} randomly oriented orbit of \psr with respect to Earth, and where a WD companion would have a mass between $0.9-1.35\;M_{\odot}$.
However, the masses of the first-born NSs in DNS systems that have been measured thus far fall between $1.30-1.60\;M_{\sun}$ \citep{of16}.
Hence, assuming this would also apply to the potential 
NS companion of \psr, we obtain an estimate for the probability of this system being a DNS system to be about 24\%. 
Assuming a most likely WD mass in the interval $1.0-1.35\;M_{\odot}$ (see Section~\ref{subsec:Porb_ecc} and Fig.~\ref{fig:heatmap+histogram})
the probability of \psr being a DNS system converges to a final value of about $\sim$32\%.
This value changes by less than 0.5\% when considering a pulsar mass between $1.15-1.45\;M_{\sun}$. 

This is a remarkable increase in probability (compared to that from the mass function alone) and suddenly makes a DNS system a much more likely scenario. Moreover, there are further constraints from the orbital parameters which we now discuss.

\section{Population synthesis}\label{sec:popsyn}

\begin{figure}
\includegraphics[width=\columnwidth]{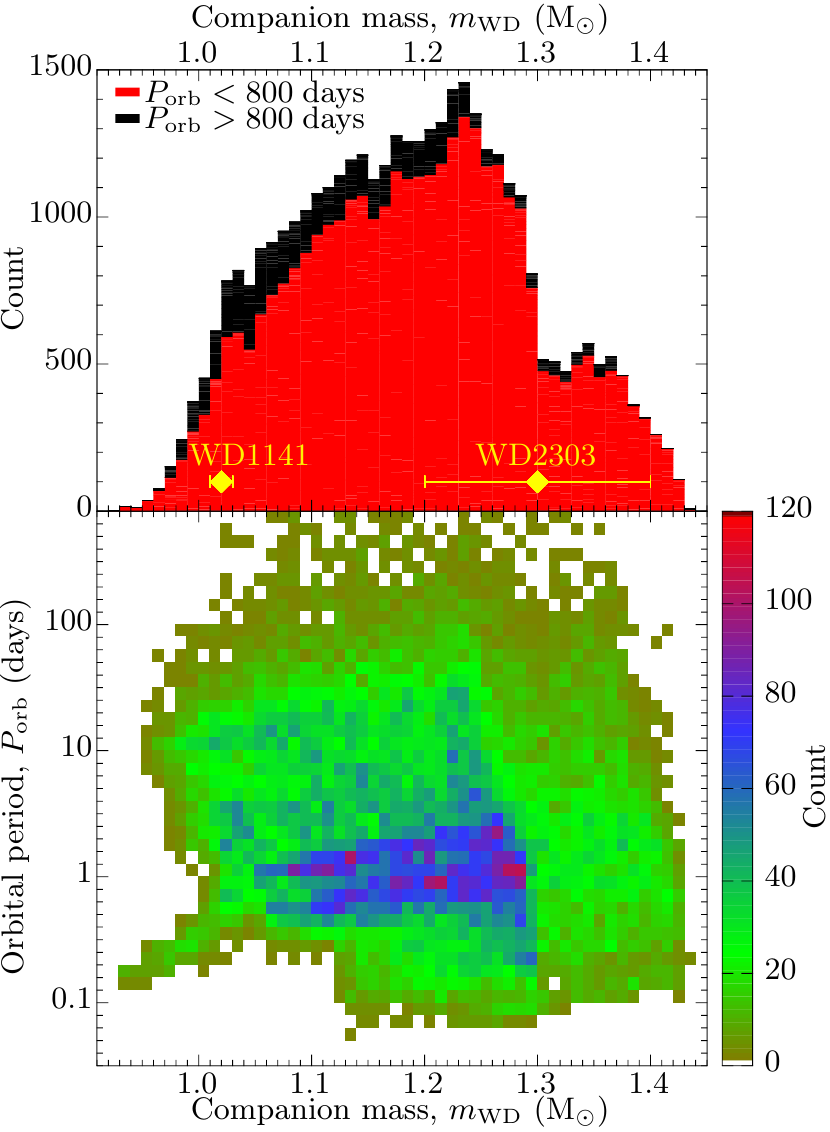}
\caption{
\textit{Bottom plot}: 
Heatmap showing WD mass and orbital period for a population of WDNS binaries with an orbital period of less than 800\,days obtained from our population synthesis simulations. The shown orbital period is the initial value right after the young NS has formed. The colour coding reflects the number of counts in each pixel. 
\textit{Top plot}: A corresponding histogram of the WD mass distributions with orbital periods below (red) and above (black) 800\,days.}
\label{fig:heatmap+histogram}
\end{figure}

To distinguish further between a WDNS and a DNS nature of the \psr system, we have taken advantage of population synthesis and simulated a large number of WDNS and DNS systems. 
We use \OurCode \citep{ktl+18}, an updated version of the population synthesis code which was previously applied to WDNS systems by \citet{ts00} and DNS systems by \citet{vt03}.
At Milky Way metallicity, \OurCode interpolates the stellar models of \citet{bdc+11} and includes new calculations of their extension to lower masses. In the applied mass range, a simple power-law of $-2.7$ is assumed for the initial mass function.
The initial binary separations are limited between $10$ and $10^4\;R_{\odot}$ and follow a flat distribution of the logarithmic orbital period. As most interacting binaries circularise before and during mass transfer, we apply initial circular orbits.

To produce WDNS and DNS systems from progenitor stars at Milky Way metallicity, we find that the ZAMS mass range of primary and secondary stars is between 5 and $35\;M_{\odot}$. The binaries which evolve to a system like \psr are well within this range. As the mass range is restricted and the initial mass function favours low-mass stars, we simulated 50~million systems to produce Figs.~\ref{fig:heatmap+histogram} and \ref{fig:sim}. All other parameters are like those in the default parameter set of \citet{ktl+18}. This especially includes the assumption of rather inefficient mass transfer, a common envelope treatment with envelope structure information obtained from the detailed stellar models, and a SN kick distribution depending on the amount of mass stripping of the progenitor star via binary interactions prior to its explosion. In the analysis, we focus on the main population with orbital periods less than $800\,{\rm days}$ at the formation of the young NS, since wider systems would likely have avoided any binary interactions.

\subsection{Orbital period and eccentricity}
\label{subsec:Porb_ecc}
\begin{figure}
\includegraphics[width=\columnwidth]{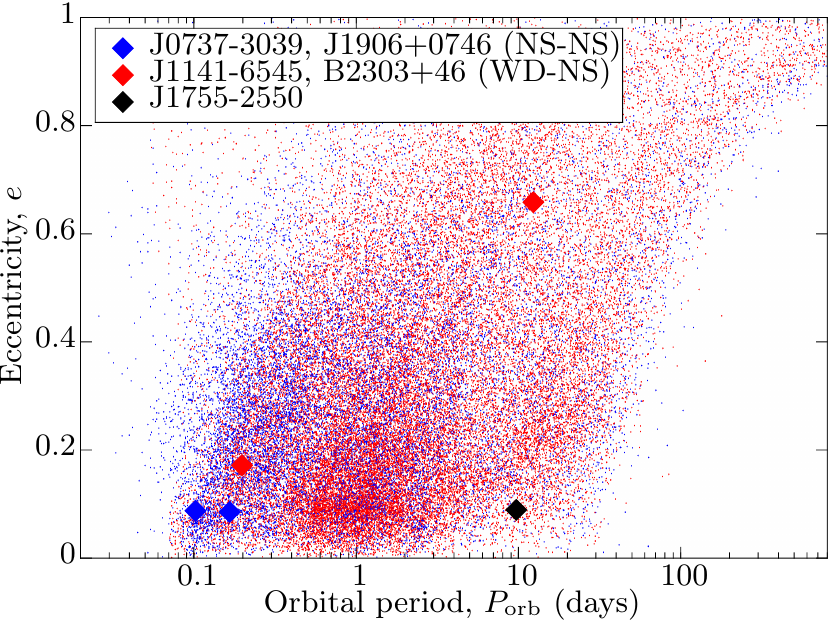} 
\includegraphics[width=\columnwidth]{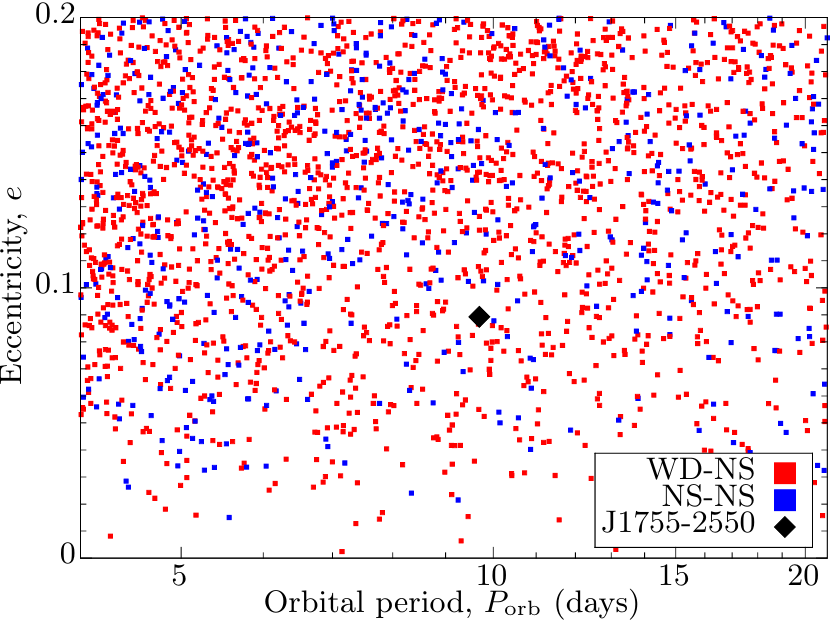}
\caption{{\em Top plot}: Orbital eccentricity ($e$) versus orbital period ($P_{\rm orb}$) of the simulated WDNS systems (red squares) and DNS  systems (blue squares) obtained from our population synthesis. The location of \psr is marked with a black diamond. The colored diamonds are the other four known young pulsars in binary systems: PSRs~J0737$-$3039B \citep{ksm+06}, J1906$+$0746 \citep{lsf+06}, J1141$-$6545 \citep{bbv08} and B2303$+$46 \citep{std85}.
{\em Bottom plot}: A zoom-in of the top plot. Both simulated WDNS and DNS systems are seen to populate the area close to \psr.}
\label{fig:sim}
\end{figure}

Fig.~\ref{fig:heatmap+histogram} shows the distribution of our simulated WDNS systems in the WD mass--orbital period plane and a histogram of the WD masses in these simulated WDNS systems. We note that indeed the majority ($>95$\%) of the simulated WDs have masses between $1.0-1.35\;M_{\odot}$. For comparison, we have plotted the WD masses of the only two known WDNS systems, PSRs~J1141$-$6545 and B2303+46. 

Fig.~\ref{fig:sim} shows the orbital period and eccentricity distribution for all simulated WDNS and DNS systems. This plot shows the orbital parameters at the birth of the double degenerate systems. However, the binaries which are born in relatively close orbits ($P_{\rm orb}\la 1\;{\rm day}$, and especially those which are eccentric) will experience gravitational damping and evolve to shorter periods and more circular configurations. For example, the tight binary PSR~J1141$-$6545 has a measured orbital period decay of  $\dot{P}_{\rm orb}=-4.0\times10^{-13}$ \citep{bbv08}.

A zoom-in in the region near the orbital parameters of \psr is shown in the bottom panel of Fig.~\ref{fig:sim}.
We notice that we can reproduce the location of \psr, and its relatively large value of $P_{\rm orb}\simeq 9.7$\,d and small value of $e=0.09$, for both WDNS and DNS systems. 
Based on our population synthesis modelling, we therefore conclude that from this information alone, we cannot distinguish between the two different possibilities (WDNS vs DNS) for the nature of \psr. 

The eccentricity of \psr is quite small (especially given its relatively large orbital period), compared to our simulated systems in general. This could be an indication of a small kick imparted on the NS in the SN explosion, or simply a kick direction which favors small post-SN eccentricities.
For comparison, the eccentricities of PSRs~J1141$-$6545 and B2303$+$46 (the two systems
where we know the companions are WDs) are $e=0.17$ and $e=0.66$, respectively. Interestingly,
the orbital eccentricities at birth for the other two young pulsars (PSR~J1906$+$0746, which could be in a DNS, and PSR~J0737$-$3039B, which certainly is in
a DNS) are 0.085 and 0.11 respectively \citep{lfl+06}. These are quite similar to the orbital eccentricity of \psr.

If \psr is a DNS system, recent simulations of the kinematic effects of the second SN explosion in this system \citep{tkf+17} indicate that the kick velocity was most likely less than $100\;{\rm km\,s}^{-1}$ (although a small tail of larger kick solutions exists). This is similar to their
findings for PSR~J0737$-$3039A/B and PSR~J1906+0746 (which nonetheless has a much larger high-velocity tail as it is not as well 
constrained). The former system is strongly constrained to kicks smaller than $70 \, \rm km \, s^{-1}$ if its very small proper motion is taken into account.

\subsection{Relative formation rate of WDNS versus DNS systems} 
From our simulations, we find that the relative formation rate of WDNS and DNS systems are quite similar: 70\% and 30\%, respectively. A thorough investigation of this ratio, and how it depends on initial parameters and various physical assumptions, is beyond the scope of this paper. However, we do notice that a roughly equal formation rate of these two subpopulations of binary pulsars systems does not seem unreasonable, given that (besides \psr) two systems of each kind have been discovered so far.

\subsection{On the possibility of a MS companion star}
\label{subsec:popsyn_MS}
Using the population synthesis code \OurCode \citep{ktl+18}, we also simulated a large population of NS--MS systems based on 50~million initial ZAMS binaries. Here, we considered MS companion stars within the mass interval of $0.4$ to $1.0\;M_{\odot}$, given the constraint from pulsar timing on the minimum mass of the companion star (Table~\ref{tab:timing}) and the lack of an optical counterpart (Section~\ref{subsec:optical}). Of these systems, only less than 2~per~cent have eccentricities below 0.10 (and $<4$~per~cent have $e<0.15$) right after the SN explosion. Subsequent long-term tidal interactions could in principle help to circularise more high-eccentricity systems. However, the circularisation timescale, $\tau _{\rm circ}$ \citep[e.g.][and references therein]{cc97} scales with orbital period and stellar radius to large powers (up to $\tau_{\rm circ}\propto P_{\rm orb}^7\,R^{-9}$ in the case of stars with convective cores and radiative envelopes) and thus these systems with relatively wide orbits will only start to circularise efficiently once the companion star becomes a red giant star. Therefore, they will not produce NS-MS binaries with $e\sim 0.1$.

Although population synthesis always comes with uncertainties based on the applied input physics \citep[e.g. the common-envelope phase prior to the SN creating the NS][]{tv06}, we find it reasonable to state that only relatively few NS-MS systems end up in orbits with small eccentricities and orbital periods somewhat resembling that of \psr. We do find many NS-MS systems with orbital periods of less than about 5~days (the progenitors of many low-mass X-ray binaries and short-orbital period binary MSPs). A thorough analysis of NS-MS systems, however, is beyond the scope of this paper. 

In contrast, the NS in systems with a compact object companion (i.e. WDNS or DNS systems) and an eccentricity and period like \psr (shown in Fig.~\ref{fig:sim}) are more common by more than an order of magnitude (compared to NS-MS binaries) according to our population synthesis simulations. Thus based on our binary modelling alone, although we cannot exclude a MS companion star to \psr, we find it most likely that the companion star is a compact object.

\subsection{Merging WDNS systems and Ca-rich SNe}
\begin{figure}
\includegraphics[width=\columnwidth]{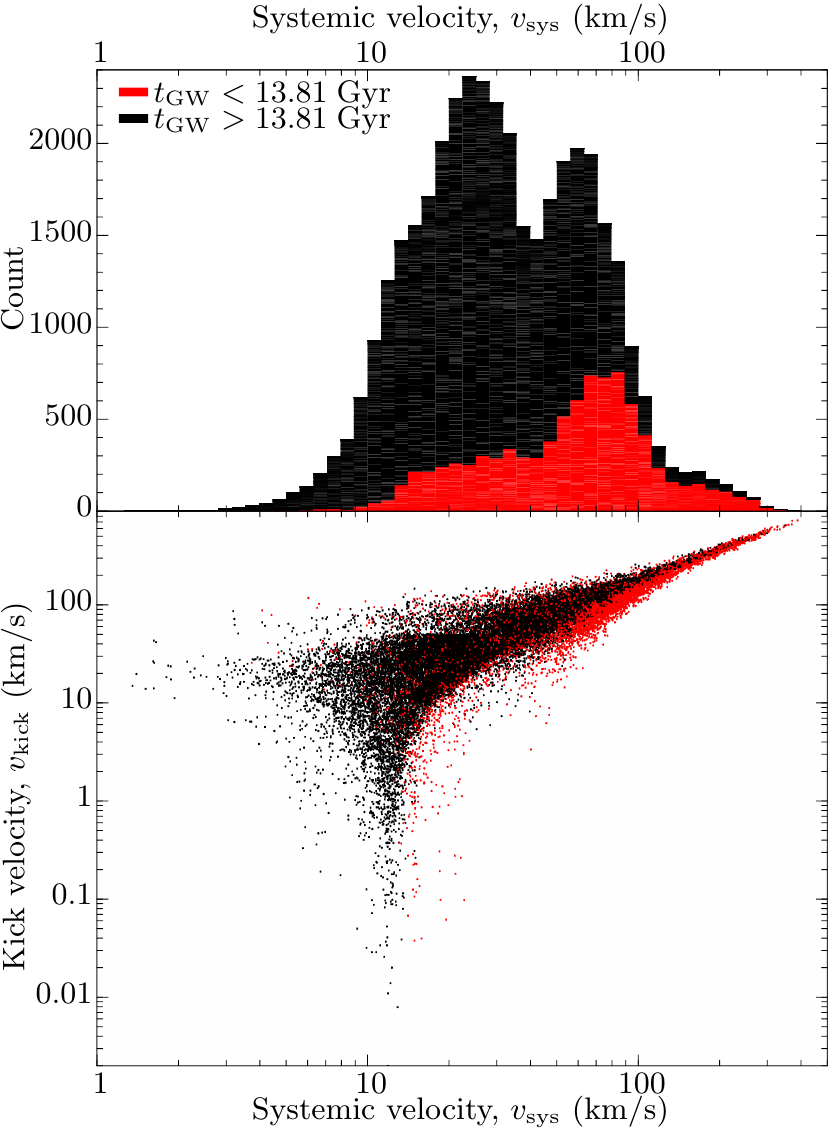}
\caption{
\textit{Lower panel}: The relation between applied NS kick velocities and resulting 3D systemic velocities of all simulated WDNS binaries. \textit{Upper panel}: A histogram of 3D systemic velocities from our simulated population of WDNS systems. The red and black distributions are for WDNS systems which merge via gravitational wave (GW) radiation before and after a Hubble time, respectively.}
\label{fig:v_sys}
\end{figure}

It has been suggested that the merger event of a massive WD and a NS might produce a Ca-rich SN \citep{met12}. These observed dim SNe (or transients) are often found at large offset distances from their associated host galaxies \citep{kkg+12,fol15}. For example, it was shown by \citet{llj+16} that about 1/3 of these Ca-rich SNe have offsets $>20\;{\rm kpc}$. Hence, to explain this extreme offset distribution many scenarios have been proposed, including unusual formation sites such as globular clusters or dwarf satellite galaxies, which are difficult to detect. \citet{mmt+17} argued that some of the Ca-rich gap transients might be related to explosions of ultra-stripped stars.

To probe whether such offset Ca-rich SNe could originate from our simulated WDNS systems escaping their birth sites, we plot in Fig.~\ref{fig:v_sys} the resulting 3D systemic velocities of our simulated population of WDNS binaries. It is seen that those WDNS systems which merge within a Hubble time typically have velocities of less than $100\;{\rm km\,s}^{-1}$, although a small high-velocity tail extends up to $350\;{\rm km\,s}^{-1}$. 

These resulting systemic velocities depend, of course, on the applied NS kick distribution. Here, the applied NS kick distribution is taken from \citet{ktl+18}, which reflects that many ultra-stripped SNe result in newborn NSs with small kicks \citep{tlp15,tkf+17}.
In the lower panel of Fig.~\ref{fig:v_sys}, we show the correlation between the applied NS kick velocities and the resulting systemic velocities of our simulated WDNS systems.

We conclude that our simulated WDNS systems are not able to escape the gravitational potential of somewhat massive host galaxies (like our Milky Way) if they originate from a disk population. Mergers of NSs and massive WDs, however, can also be produced from NSWD systems such as PSR~J1952+2630 \citep{ltk+14}, i.e. binaries with a recycled pulsar and thus where the NS forms before the WD. A kinematic investigation of those systems is beyond the scope of this paper. Finally, we note that the similar peak luminosity of Ca-rich SNe to those of SNe~Ib has led to the suggestion that Ca-rich SNe may also arise from the core collapse of massive stars \citep[e.g.][]{glf+17}. How to account for the often observed large offsets of Ca-rich SNe with respect to their host galaxies remains to be explained in this formation model --- unless multiple formation paths for Ca-rich SNe are possible.

Regardless of \psr being a WDNS or a DNS system, its orbital period is much too large to produce a Galactic merger event. However, the kinematics of both populations is dominated by the SN explosion producing the last-formed compact object, and thus we predict a proper motion of \psr consistent with a 3D systemic velocity of less than $100\;{\rm km\,s}^{-1}$.

\section{Observable clues}\label{sec:discussions}
\subsection{Optical search of a first-formed WD}
\label{subsec:optical}
\begin{figure}
\centering
\includegraphics[width=3.2in]{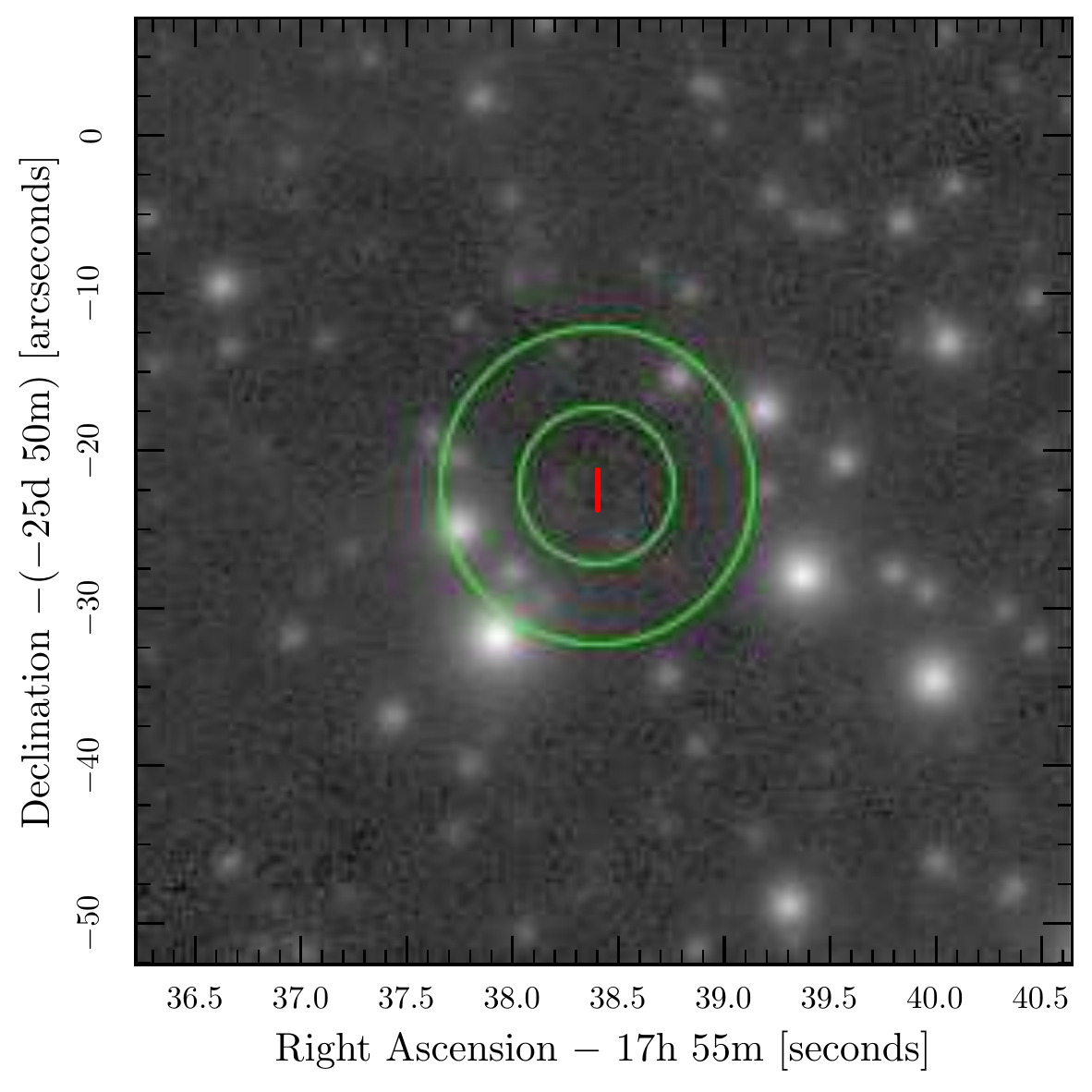} 
\caption{The stacked Pan-STARRS data in $y-$band. The circles indicate 5 and 10~arcsec radius around the position of \psr from radio timing. The red rectangle shows the uncertainty of position from radio timing.} 
\label{fig:opt}
\end{figure}

\begin{figure}
\centering
\includegraphics[width=3.2in]{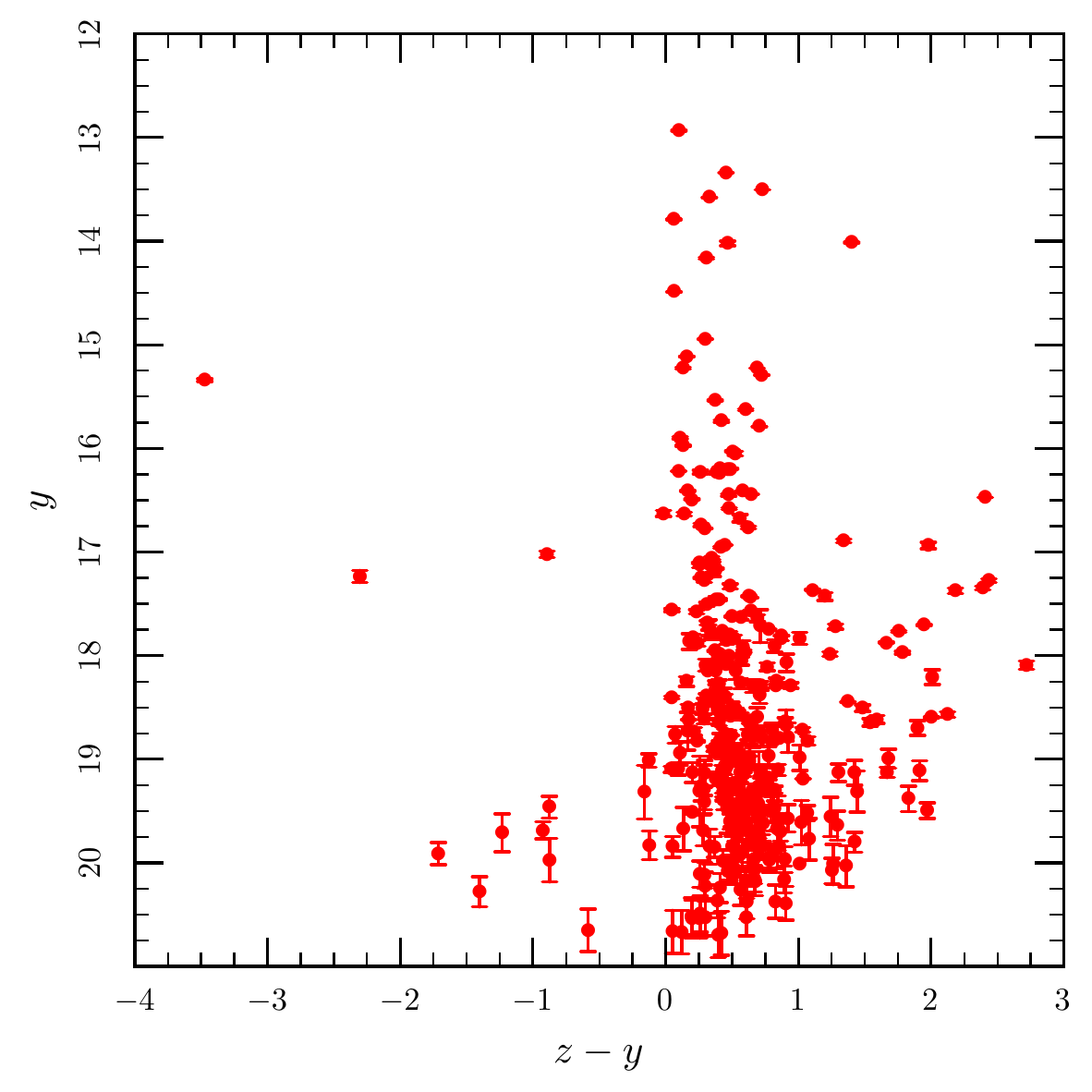}
\caption{Colour-Magnitude Diagram of objects in the Pan-STARRS catalogue \citep{2016arXiv161205560C} within one arcminute of the position of \psr from radio timing.  The 5-$\sigma$ detection limit is $z=22.3$ and $y=21.3$.}
\label{fig:cmd}
\end{figure}

Any optical detection of the companion would provide definitive evidence for the WD argument in the case of \psr. 
The deepest archival data we found covering the vicinity of \psr is from the pan-STARRS survey \citep{cha06,kai10}. There are five available bands ($g, r, i, z, y$) in total. 
We performed point spread function (PSF) photometry of each filter using \textsc{daophot}. \citet{2016arXiv161205242M} find that the mean astrometric deviation relative to the GAIA catalogue is about 5~milliarcseconds in this region of the sky, and furthermore the 
astrometry precision of the translation between the pulsar timing frame of reference and the GAIA frame of reference is likely to be even smaller \citep{2017MNRAS.469..425W}, so the dominant positional uncertainties are those from the timing measurements themselves (Tab.~\ref{tab:timing}) and are about two arcseconds.
Since \psr has a very low Galactic latitude, absorption is likely significant along the line-of-sight. We thus focus our analysis on the $y-$band (around 1 micrometer) data where the amount of absorption is the least among the pan-STARRS filters (Fig.~\ref{fig:opt}). We find no object at the position of \psr down to a detection limit of $z=22.3$ and $y=21.3$ from the stacked image (Fig.~ \ref{fig:cmd}).

\begin{figure}
\includegraphics[width=\columnwidth]{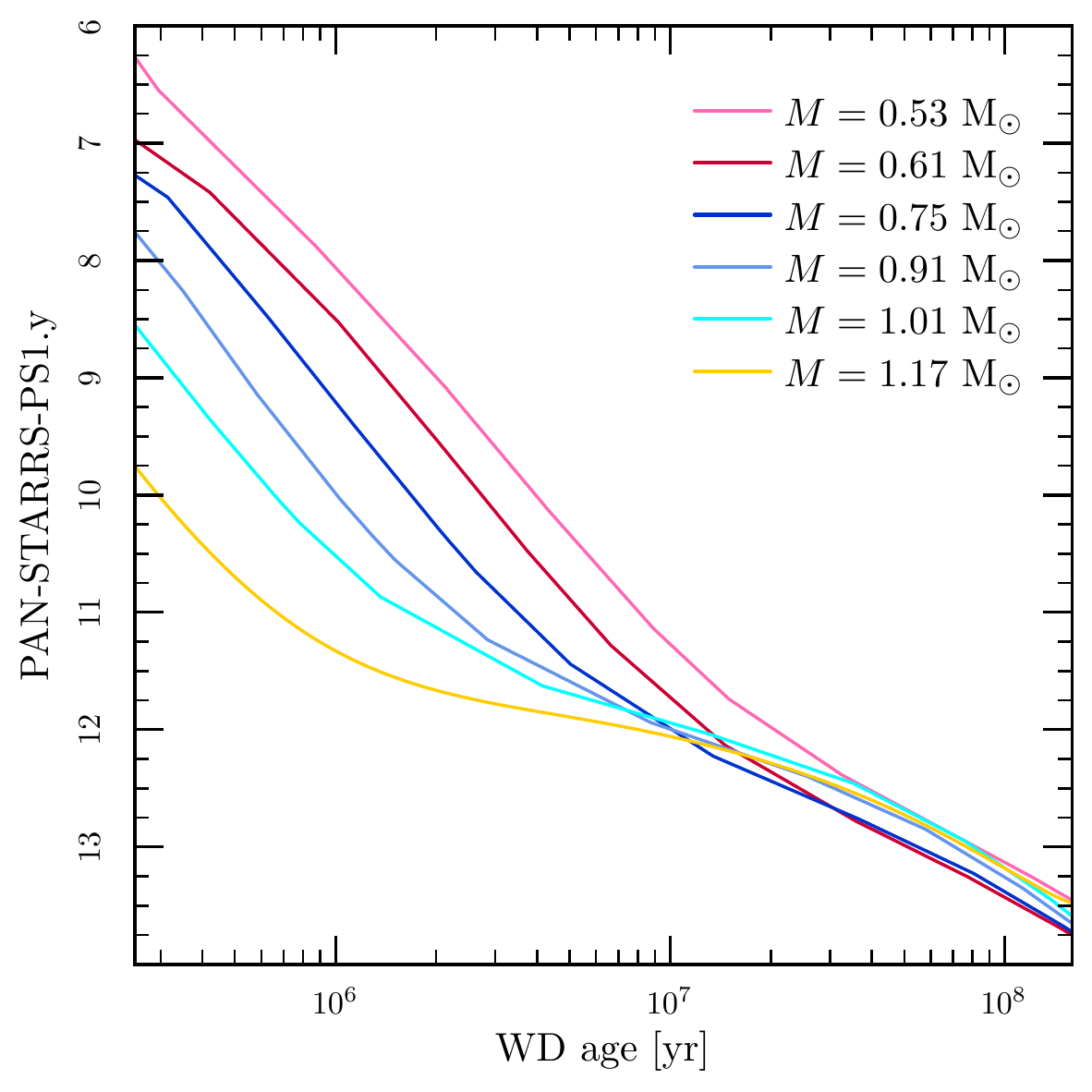} 
\includegraphics[width=\columnwidth]{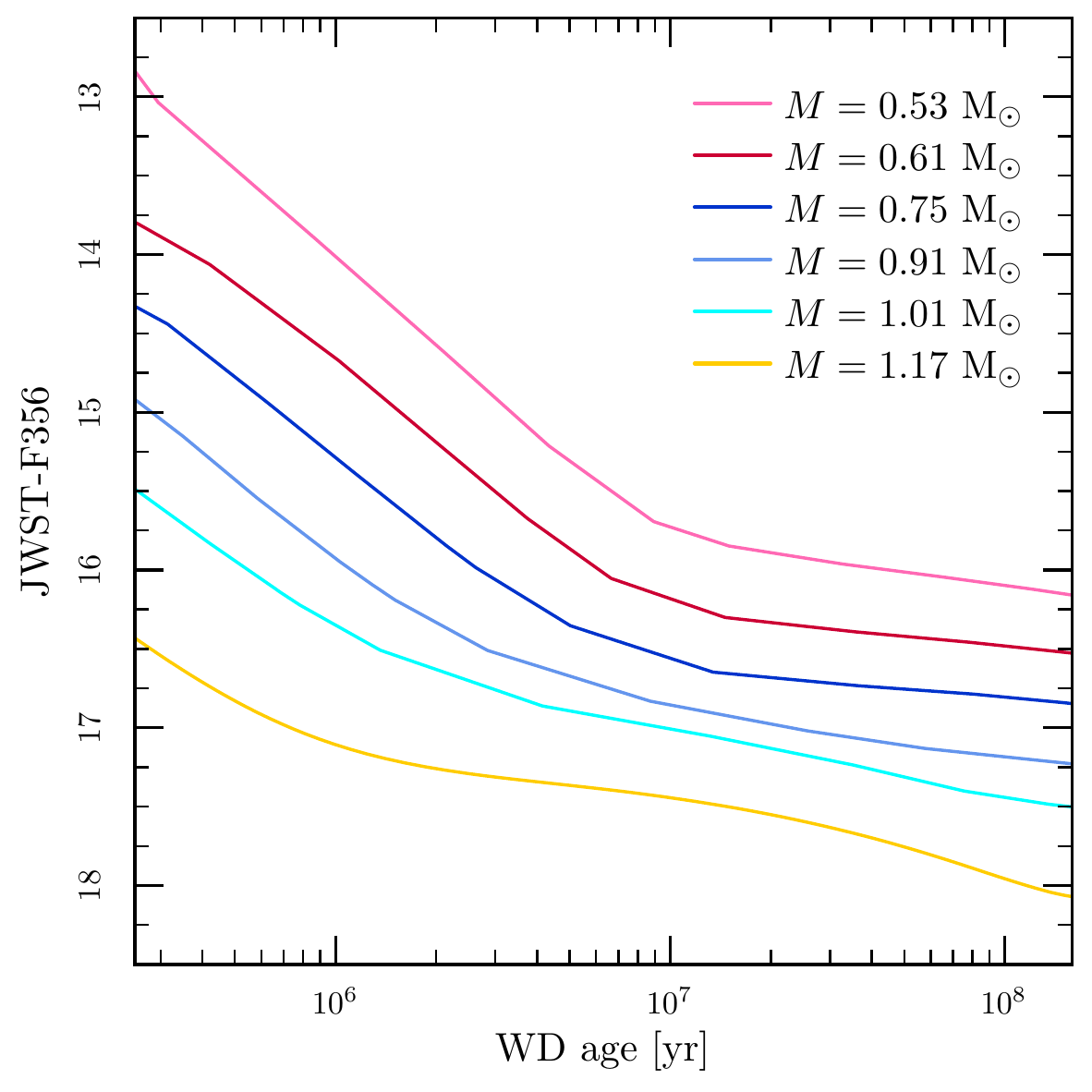}
\caption{ White dwarf cooling models for 6 different WD masses. {\em Top plot}: Absolute magnitude vs WD age in the pan-STARRS $y-$band.
{\em Bottom plot}: Absolute magnitude vs WD age in JWST F356W filter.}
\label{fig:wdcooling}
\end{figure}

Using the measured DM of 751\,cm$^{-3}$pc and the \citet{cl02} model (hereafter NE2001) of free electron distribution in the Milky Way, we obtain a DM-derived distance of 10.3\,kpc. However, a recent electron density model \citep[][hereafter YMW16]{YMW16} puts the DM-derived distance of \psr much closer, at 4.9\,kpc.

In Fig.~\ref{fig:wdcooling} we plot theoretical cooling curves for WD masses ranging from 0.53 to 1.17\,M$_{\odot}$ calculated with MESA \citep{pca+13} and the DA stellar atmospheres of \citet{2006AJ....132.1221H}, \citet{2006ApJ...651L.137K} and \citet{2011ApJ...730..128T}\footnote{http://www.astro.umontreal.ca/$\sim$bergeron/CoolingModels}. The top panel shows the absolute magnitude in the pan-STARRS y-band. The uncertainty in the distance of \psr makes it hard to quantify a theoretical apparent magnitude. Assuming a minimum distance of 4.9\,kpc and no extinction, we have to add 13.46 magnitudes (the distance modulus corresponding to 4.9\,kpc) to the values on the y-axis. As mentioned, \psr is located very close to the Galactic plane and even in the y-band we have a minimum extinction of $\sim$5 magnitudes \citep{1998ApJ...500..525S,2011ApJ...737..103S}\footnote{http://irsa.ipac.caltech.edu/applications/DUST/}. These sum to an apparent magnitude of 25-30 for a 1\,Myr-old WD (roughly the spin-down age of the pulsar) depending on its mass, which is well beyond the limit of the sensitivity of the archival pan-STARRS data, and thus our non-detection is not conclusive. In the lower panel of Fig.~\ref{fig:wdcooling}, we show the same cooling models if we were to observe with the F356W filter of the James Webb Space Telescope (JWST). WDs are fainter in this redder filter. However, the absorption in this band is much lower, less than a magnitude at a distance of $\sim$5\,kpc. A constraining optical detection with the JWST could be possible although still challenging.

In Sections~\ref{subsec:MS-companion} and~\ref{subsec:popsyn_MS}, we discussed the scenario of the companion being a MS star.  In particular eclipses and changes in the orbital period are often seen in pulsars with MS companions.  Furthermore, population synthesis indicates that it is difficult to form a binary with eccentricity as low as the \psr system. The pan-STARRS data can provide some additional clues. Although a low-mass MS star would have been too dim to be detected, a MS star with a mass of $\sim$0.9\,$M_{\odot}$ would have an absolute magnitude of about 3 in the pan-STARRS $y-$band. Therefore, if \psr was at the minimum distance of $\sim$5\,kpc, then a MS star more massive than 0.9\,$M_{\odot}$ would have an apparent magnitude brighter than 21, and would have been detected in the pan-STARRS data studied here. This means that if we knew for certain that \psr was at $\sim$5\,kpc, we could exclude a MS companion more massive than 0.9\,$M_{\odot}$. However, at a distance of ten or more kpc, extinction would make it impossible to detect even a much higher mass MS star. A post-MS star would be even brighter. Again, the uncertainty on the distance makes it difficult to draw conclusions. There is no evidence of an MS or post-MS companion more massive than 0.9\,$M_{\odot}$, but its presence cannot be excluded by the optical data unless the \psr is about five or fewer kiloparsecs away.

\begin{table*}
    \centering
  \caption{Search-mode observations used to search for radio pulsations from the companion of PSR J1755$-$2550.}
  \label{tab:search_mode_data}
\begin{tabular}{ccccccccccc}
\hline
Telescope  &   Date    & Central  & Bandwidth &  Gain & System & Sampling & Length & Notes & $S_{\rm limit}$  & $L_{\rm limit}$\\
  &       & Freq. (MHz) & (MHz) & (K\,Jy$^{-1}$) & Temp. (K) & Time ($\upmu$s) &  (s) &  & (mJy) & (mJy\,kpc$^{2}$)\\
\hline
Parkes     & 04/04/2013 & 1352 & 340 & 0.74 & 30.6 & 64.00& 4300 & & 0.07 & 1.7$-$7.4\\
Effelsberg & 30/08/2015 & 1347 & 200 & \multirow{2}{*}{1.37} & \multirow{2}{*}{22} &61.44 & 7150 & & 0.02 & 0.5$-$2.1\\
Effelsberg & 31/08/2015 & 1347 & 200 & & &248.32 & 3590 & Severe RFI & $-$\\
Effelsberg & 01/09/2015 & 2627 & 200 & \multirow{2}{*}{1.50}& \multirow{2}{*}{17} & 248.32 & 6580 & Severe RFI & $-$ \\
Effelsberg & 02/09/2015 & 2627 & 200 & & &248.32 & 9930 & Severe RFI & $-$ \\
Effelsberg & 04/09/2015 & 4837 & 200 & \multirow{2}{*}{1.55}& \multirow{2}{*}{27} & 61.44 & 8460 & & 0.02 & 0.5$-$2.1\\
Effelsberg & 05/09/2015 & 4800 & 125 & & & 245.76 & 8310 & & 0.02 & 0.5$-$2.1\\
\hline
 \end{tabular}
\end{table*}
\subsection{Search for radio pulsation from the potential first-formed NS}
{ If \psr is indeed a DNS system, we know from observations of the first-formed (recycled) NSs in other DNS systems, that there is roughly a one-in-three chance that it would be beaming in our direction \citep{kxl+98}, yielding an estimated 10\% probability of detecting radio pulsations from the companion star of \psr (given our estimation of a $\sim\!32$\% chance of \psr being a DNS system, cf. Section~\ref{subsec:mass-function}). If detected, it would make \psr the only other observed double pulsar system apart from PSR~J0737$-$3039 \citep{bdp+03,lbk+04}. 

To look for radio pulsations from the potential first-formed NS companion, we searched the 
seven observations available to us (Table \ref{tab:search_mode_data}). These consist of one Parkes scan in filterbank-mode taken in 2013 and six baseband observations from the Effelsberg radio telescope in 2015, covering multiple frequency bands (1.3, 2.6, and 4.8\,GHz). Three observations, however, were severely affected by RFI, which made the data unusable for search purposes.
We were thus left with two 1.3-GHz and two 4.8-GHz observations, each of which was processed as follows.

The \texttt{PRESTO}\footnote{http://www.cv.nrao.edu/$\sim$sransom/presto} routine \texttt{rfifind} was used to identify RFI in the time and frequency domain and the resultant mask was employed in subsequent processing. 
The observing band was then de-dispersed using the \texttt{prepsubband} routine.
In the case of the two 4.8-GHz observations, de-dispersion was done only once, at the nominal measured DM of the pulsar ($751\pm3$~cm$^{-3}\,$pc). Indeed, at such a high observing frequency, the possible pulse drift across the observing band due to an incorrect DM is, within the uncertainty, negligible ($\ll$ 1\,ms).
On the contrary, at 1.3\,GHz, an error of $\sim0.75$~cm$^{-3}\,$pc is sufficient to cause a pulse drift of 1 ms across an observing band of 400\,MHz. This would result in a significant smearing of the signal in the case of an MSP. For this reason, the Parkes and Effelsberg 1.3-GHz observations were de-dispersed at multiple DM values, covering the 2-$\sigma$ uncertainty range (745$-$757~cm$^{-3}$\,pc), with steps of 0.25 and 0.50\,cm$^{-3}$\,pc, respectively. This guaranteed a maximum pulse drift of a fraction of a ms across the band, for the best DM trial, in both observations.
After de-dispersing and taking the previously-made mask into account, the \texttt{prepsubband} routine also summed the frequency channels and referred each time sample to the Solar System barycentre (SSB), thus producing RFI-free, barycentered, de-dispersed time series.

To maximize our search sensitivity, we completely removed the effect of the orbital motion, in order to make the putative companion pulsar appear as if it were isolated. This was achieved utilizing a code that has been developed and previously used for the search of the possible pulsar companion of PSR J0453+1559 \citep{msf+15}. The code recalculates the time stamp of each sample of the time series by subtracting the R{\o}mer delay due to the companion orbital motion. The time series is then made uniformly-sampled again through a linear interpolation.

The R{\o}mer delay associated with the companion depends on the characteristics of the companion orbit. The latter are all known with the exception of the projected semi-major axis, $x_{\rm c}$. This, however, is directly related to the projected semi-major axis of the pulsar orbit, through the equation $x_{\rm c} = q \, x_{\rm p}$, where $q = m_{\rm p} / m_{\rm c}$ is the ratio between the mass of the pulsar ($m_{\rm p}$) and that of the companion ($m_{\rm c}$). 

Because $q$ is currently unknown, the parameter was searched in the range $0.32 - 3.125$. Such an interval more than covers all the plausible values for the possible mass range of NSs. Indeed, assuming a minimum companion mass of 0.8\,M$_\odot$ (an even more conservative value than the 0.9\,M$_\odot$ 
derived from our population synthesis) $q \leq 3.125$ for $m_{\rm p} \leq 2.5\,{\rm M}_\odot$; on the other hand, by exchanging the roles, (thus assuming $m_{\rm p} \geq 0.8\,{\rm M}_\odot$ and $m_{\rm c} \leq 2.5\,{\rm M}_\odot$) we have $q\geq0.32$.
The choice of the step size, $\Delta q$, was also crucial to minimize the total computational time and to avoid the production of an unnecessarily large number of candidates, without degrading our sensitivity.
The criterion used was the following: for each observation, we chose 
$\Delta q$ such that, in the case of the best trial value, the residual orbital modulation would cause a maximum pulse drift, seen within the observation length, of less than 0.25 ms (i.e. less than a quarter of spin period of the fastest possible recycled MSPs allowed by theoretical models).
This depends on the particular orbital phase range spanned by the companion in the considered observation, which can be easily computed through the pulsar ephemeris.
The de-modulated time series was then searched with the \texttt{accelsearch} routine, summing up to 8 harmonics and with no acceleration.

The search, done for each $q$ and DM trial, produced several hundred thousands of candidates, which were successively greatly reduced in number by sifting algorithms. The final few thousands of candidates were folded using the corresponding de-dispersed time series and the resulting plots were inspected  visually. For the 79 most convincing candidates we also folded the corresponding original filterbank file, to examine the signal in the frequency domain. The vast majority of the signals turned out to be very narrow-band RFI that were not detected during the masking procedure. None of the remaining candidate signals could be clearly ascribed to a pulsar. We conclude that no signal coming from the companion was detected. 

One effect that could lead to the non-detection is the degradation in detectability due to scattering broadening. The \citet{cl02} electronic density model predicts a scattering timescale of $\sim$66\,ms along the line-of-sight of \psr at an observing frequency of 1\,GHz. 
As mentioned in Section~\ref{sec:profile}, we measure a much smaller characteristic scattering timescale of  $\sim$13.5(14)\,ms at 1\,GHz. Indeed, the profile of \psr (see Fig.~\ref{fig:pol}) does not visually appear very scattered.
Nonetheless, this amount of scattering will surely prevent any MSP companion from being detected at 1.3\,GHz. The scattering broadening should be less severe at 2.6\,GHz, but at the same time, the pulsar flux density is also likely to get smaller.

Notwithstanding, our non-detections can be used to estimate limits on the flux density, at the different frequencies, by applying the radiometer equation and assuming a pulse duty cycle of 5\,\%. In turn, we can infer a luminosity limit of the putative companion. We caution that any derived luminosities are dependent on our knowledge of the pulsar distances, which is not well determined in the case of \psr. In Table \ref{tab:search_mode_data}, we quote the luminosity limit using both the NE2001 and YMW16 electron density model. The NE2001 model gives a higher DM distance compared to the YMW16 model which is responsible for the more conservative (higher) luminosity limit. In either case, our luminosity limit is comparable to the lower bound of the known pulsar population \citep[see, for example, Fig.~11 in][]{ncb+15}. The existence of a weak radio pulsar companion thus cannot be entirely ruled out from our radio search.

\section{Conclusion} \label{sec:conclusion}
We have observed the recently discovered radio pulsar \psr \citep{ncb+15} over a timespan of 2.6~yr using the Lovell Radio Telescope, the Parkes Radio Telescope and the Effelsberg Radio Telescope.
We find that this pulsar has a large spin-period derivative of $2.4\times10^{-15}$. 
The combination of this value with a slow spin period ($P=315\;{\rm ms}$) and a non-circular orbit ($e=0.089$), identifies \psr as being a non-recycled radio pulsar. From its mass function, we have deduced that \psr is a member of a binary system with a companion star in the mass range of $0.4-2.0\;M_{\odot}$ at the 95\% C.L. The nature of the companion star is most likely restricted to the following three possibilities: a MS star, a WD or a NS.
From a comparison to other radio pulsars with hydrogen-rich companions, we find it unlikely that this system has a MS star companion. We thus propose that \psr is the second-formed object in a double compact object binary.

Applying population synthesis modelling, we find that the chances of \psr being a WDNS system or a DNS system are roughly equal. Our population synthesis also predicts a minimum companion mass of 0.90\,$M_{\odot}$ and typically a systemic velocity of less than $100\;{\rm km\,s}^{-1}$. We conclude that \psr could very well be a DNS system and we estimate in total a $\sim$~10$\,\%$ chance of detecting its companion star as a recycled radio pulsar (taking into account both the uncertainty of the nature of the companion star and the beaming fraction in case it is a DNS system). Our effort of searching for the plausible radio pulsar companion returns no detection and our luminosity limit is comparable to the lower bound of the known pulsar population. We also attempted to look for signs of a WD companion in archival Pan-STARRS data, but find no object at the position of \psr down to a detection limit of $z=22.3$ and $y=21.3$.

Whether \psr hosts a WD or a NS companion star, it will only be the third ever known WDNS, or the third ever known DNS system where we observe the second-formed NS, and thus it represents a rare subpopulation of binary pulsars in either case.

\section{acknowledgements}
The Parkes Observatory is part of the Australia Telescope National Facility, which is funded by the Commonwealth of Australia for operation as a National Facility managed by CSIRO. A part of this work is based on observations with the 100-m telescope of the MPIfR (Max-Planck-Institut f\"{u}r Radioastronomie) at Effelsberg. Pulsar research at JBCA and access to the Lovell Telescope is supported by a Consolidated Grant from the UK Science and Technology Facilities Council (STFC).
CN is supported by an NSERC Discovery Grant and Discovery Accelerator Supplement and by the Canadian Institute for Advanced Research. MUK acknowledges financial support by the DFG Grant: TA~964/1-1 awarded to TMT. AR and PCCF gratefully acknowledge financial support by the European Research Council for the ERC Starting grant BEACON under contract no. 279702, and continuing support from the Max Planck Society.
AR is member of the International Max Planck research school for Astronomy and Astrophysics at the Universities of Bonn and Cologne and acknowledges partial support through the Bonn-Cologne Graduate School of Physics and Astronomy.  IC and JH are supported by an NSERC Discovery Grant. The Pan-STARRS1 Surveys (PS1) and the PS1 public science archive have been made possible through contributions by the Institute for Astronomy, the University of Hawaii, the Pan-STARRS Project Office, the Max-Planck Society and its participating institutes, the Max Planck Institute for Astronomy, Heidelberg and the Max Planck Institute for Extraterrestrial Physics, Garching, The Johns Hopkins University, Durham University, the University of Edinburgh, the Queen's University Belfast, the Harvard-Smithsonian Center for Astrophysics, the Las Cumbres Observatory Global Telescope Network Incorporated, the National Central University of Taiwan, the Space Telescope Science Institute, the National Aeronautics and Space Administration under Grant No. NNX08AR22G issued through the Planetary Science Division of the NASA Science Mission Directorate, the National Science Foundation Grant No. AST-1238877, the University of Maryland, Eotvos Lorand University (ELTE), the Los Alamos National Laboratory, and the Gordon and Betty Moore Foundation. This research has made use of the NASA/IPAC Infrared Science Archive, which is operated by the Jet Propulsion Laboratory, California Institute of Technology, under contract with the National Aeronautics and Space Administration. The Dunlap Institute is funded by an endowment established by the David Dunlap family and the University of Toronto.
We thank Ross Church, and other participants at the NORDITA workshop on {\it The Physics of Extreme-Gravity Stars}, for discussions on Ca-rich SNe.

\bibliographystyle{mnras}
\bibliography{cherry_refs,alex_refs} 

\bsp	
\label{lastpage}
\end{document}